# Sb concentration dependent Structural and Transport properties of Polycrystalline $(Bi_{1-x}Sb_x)_2Te_3$ Mixed crystal

K. Malik[1*] and S. Mahakal[1], Diptasikha Das[2], Aritra Banerjee[3], S. Chatterjee[4], Anusree Das[4]

[1]Department of Physics, Vidyasagar Metropolitan College, 39 Sankar Ghosh Lane Kolkata-700006, India

[2]Adamas University, Barasat - Barrackpore Rd, Kolkata 700126, India

[3]Department of Physics, University of Calcutta, 92 A P C Road, Kolkata 700 009, India

[4]UGC-DAE Consortium for Scientific Research, Kolkata Centre, Sector III, LB-8, Salt Lake, Kolkata 700 106, India

## ABSTRACT

$(Bi_{1-x}Sb_x)_2Te_3$ [x=0.60, 0.65, 0.68, 0.70, 0.75 and 0.80] mixed crystals have been synthesized by solid state reaction. In depth structural, thermal, transport and electronic properties are reported. Defect and disorder play a crucial role in structural and transport behaviour. Disorder induced iso-structural phase transition is observed at x=0.70, which is supported by the structural and transport properties data. Debye temperature has been estimated from the powder diffraction data. Differential scanning calorimetry (DSC) data confirms the glass transition in the material. Low temperature resistivity data ($\rho(T)$) shows Variable range hopping mechanism whereas high temperature data follows activated behaviour. Activation energy is calculated from the semiconducting region of $\rho(T)$. Both Hall measurement and temperature dependent thermopower data (S(T)) confirms that samples are p-type in nature. Density of state effective mass has been estimated from Pisarenko relation and corroborated with resistivity data. Thermal conductivity ($\kappa$) is estimated using experimentally obtained data. Figure of Merit (ZT) of the synthesized samples are calculated using $\rho(T)$, S(T) and $\kappa$. Structural and transport properties are correlated, confirms the transition from disorder to order state. Defect and disorder are corroborated with structural and Thermoelectric properties of the synthesized samples.

---

*Author to whom correspondence should be addressed: kartickmalik@vec.ac.in     1

## I. INTRODUCTION

Thermoelectric (TE) materials are those which convert thermal to electrical energy and vice-versa [1, 2]. Efficiency of a TE material is defined by Figure of Merit, $ZT = (S^2\sigma/\kappa)T$; where S, σ, and κ are thermopower, electrical conductivity and thermal conductivity respectively [3, 4]. These are interrelated material's property. Worldwide there is resurgence to enhance the efficiency by optimizing physical properties of the TE material. Alloying is one of the potential techniques to modify material's properties [5]. Substitutional crystalline solid solution is an effective technique to modify band structure and transport properties of the TE materials. [6, 7, 8]. κ of the alloys is strongly influenced by the mass fluctuation in the crystal structure and defects. Mechanical alloying creates scattering centre which in turn affect the electrical/thermal transport properties [9, 10] of material.

Nowadays best n-type and p-type TE materials available near room temperature are $Bi_2Te_3$, $Sb_2Te_3$ based materials [11]. These binary pnictide chalcogenides are not only potential TE material but also interesting from electronic band structure point of view. Further, the mixed crystals are also belong to intriguing class of material, topological insulator (TI) [11]. These materials crystallize with layered rhombohedral crystal structure ($R\bar{3}m$). However, layered crystal structure of these material is favourable to low $\kappa \sim 2.4$ $Wm^{-1}K^{-1}$ at 300 K [15]. Chemical doping in the $Bi_2Te_3/Sb_2Te_3$ is one of the fascinating and effective tools for further enhancement of TE properties. Bi doping in $Sb_2Te_3$ results $(Bi,Sb)_2Te_3$, which is a good thermoelectric material [16,17,18]. Isoelectronic Bi and Sb form solid solution for wide range of Sb concentration [7]. $Sb_2Te_3$ and $Bi_2Te_3$ may be considered as two end members of $(Bi_{1-x}Sb_x)_2Te_3$ mixed crystal, are also promising TE material as their parent material near room temperature. $(Bi_{1-x}Sb_x)_2Te_3$ alloys form solid solution for the entire Sb concentration (x) range [11]. Recently materials are drawing attention of researcher due to their three-dimensional topological insulating properties [12, 13, 14]. D. Kong et al. have shown that TI properties are also restored for the entire composition range of $(Bi_{1-x}Sb_x)_2Te_3$ [8, 13]. Mixed crystal $(Bi_{1-x}Sb_x)_2Te_3$, an alloy of $Bi_2Te_3$ and $Sb_2Te_3$, follow the Vegard's law [19]. $(Bi_{1-x}Sb_x)_2Te_3$ are crystallized in rhombohedral phase as its end member [20,21]. Unit cells of the material are ordered as layered structure along the $C_3$ axis of the material. Each unit layer contains five atomic planes. These are bonded together by van-der-wall bond. TE materials with tetradymite family contain heavy metal and concomitantly spin-orbit coupling (SOC) plays a significant role in the electronic band structure [11, 8]. The general formula of tetradymite structure is $M_2X_3$, where M is Group V elements and X is chalcogenide [11]. However, layered structure of the material plays an important role to enhance ZT by reduction of κ [15]. Both TE and TI properties of the material also dominated by inherent defect. Weak interlayer bonding causes anti-site defect in the material. Inherent native defect in $Sb_2Te_3$ material shows p-type conductivity with carrier concentration up-to $10^{20}$ /$cm^3$ in $Sb_2Te_3$ material. Thermal variation of resistivity, $\rho(T)$ frequently shows metallic behaviour due this defect induced huge carrier concentration [16, 22]. However, Das et al. also pointed out that surface conductivity mixed with bulk conductivity rise to metallic behaviour in $Sb_2Te_3$ [23]. Reported results of $Bi_2Te_3$ show metallic behaviour like another member, although it supposes to be a band insulator [24]. The Te vacancy ($V_{Te}$) and antisite defect of Bi ($Te_{Bi}$), two prominent defects give rise to metallic behaviour in $Bi_2Te_3$ and $Sb_2Te_3$ [16, 11]. Increasing Sb doping in $Bi_2Te_3$ i.e., in $(Bi_{1-x}Sb_x)_2Te_3$ causes shift in Fermi surface ($E_F$) and bulk state completely disappear at $E_F$ for the composition $(Bi_{0.50}Sb_{0.50})_2Te_3$. However, physical properties of mixed crystal depend on the synthesis processes. Room temperature ZT as high as 3.0 x$10^{-3}$$K^{-1}$ is achieved for hot pressed $(Bi_{1-x}Sb_x)_2Te_3$ at the Sb alloying level 0.70≤x≤0.80 and shows p-type conductivity. In this processes p-type to n-type transition occurred around x=0.67 [25]. Another important route of synthesis is mechanical alloying followed by pulse discharging [26]. Efforts also have been given to decouple the bulk and surface conductivity to study the TI nature of mixed crystal [27]. Further, thermoelectric properties of low dimensional mixed crystal have been widely studied. However, study on effect of Sb concentration in the structural and transport properties of polycrystalline $(Bi_{1-x}Sb_x)_2Te_3$ are limited [25]. Amid the synthesis processes, solid state reaction process is one of the easiest ways to produce abundant amounts of sample. Reported study of synthesized samples using solid state reaction method are also limited. Furthermore, polycrystalline material with grain boundary scattering may give positive contribution in ZT. It is



noteworthy to mention that vacancy and defect along with the surface state play an important role in the structural, transport and TE properties of the mixed crystal.

In the present work, polycrystalline samples have been synthesized by solid state reaction method. In-depth structural characterizations have been performed by Rietveld refinement. Thermal variation of resistivity and thermopower measurements have been carried out and correlated with structural data. Glass transition temperature, crystallization temperature and iota of unreacted or materials at grain boundary in the matrix of the mixed crystal have been analysed by DSC (Differential Scanning calorimetry). Thermal variation of carrier concentration of the synthesized samples have been estimated. Further, effect of band structure engineering along with defect and vacancy concentration on the structural, transport properties are studied analytically. Anomalous change in TE properties is correlated with structural, DSC and carrier concentration data. In this maiden attempt iso-structural phase transition has been reported in $(Bi_{1-x}Sb_x)_2Te_3$ for x=0.30. Disorder and defects play a crucial role in transport properties and lead to iso-structural phase transition. Disorder induced iso-structural phase transition has been corroborated with transport properties data and underline physics is discussed.

## II. EXPERIMENTAL DETAILS

Polycrystalline $(Bi_{1-x}Sb_x)_2Te_3$ (0.60≤x≤0.80) samples have been synthesized by solid state reaction method. Elemental Bismuth, Antimony and Tellurium are sealed in a quartz tube under pressure of $10^{-3}$ Pa to avoid oxidation. The sealed quartz tube is annealed at temperature 1123 K for 24 hour (hr). After that, it was cooled down to 880 K in 48 hr and then sintered for 96 hr to obtain homogeneous sample. The end ingot was obtained by quenching quartz tube at liquid Nitrogen.

Structural characterization of $(Bi_{1-x}Sb_x)_2Te_3$ (for x= 0.60, 0.65, 0.68, 0.70, 0.75 and 0.80) samples have been carried out by powder x-ray diffraction (XRD) using powder x-ray diffractometer [Model: X'Pert PRO (PANalytical)] with Cu-K$_\alpha$ radiation. All the XRD measurements are performed in the range of $15^0 \leq 2\theta \leq 80^0$ in θ-2θ geometry. In-order to analyse in-depth structural parameters, Rietveld



refinement of room temperature powder diffraction data were done using MAUD (Materials Analysis Using Diffraction) software [28]. ρ(T) measurements, down to 10K have been carried out using four probe method. S(T), down to 20K were measured by differential technique [29]. DSC were performed in inert gas atmosphere at heating rate $10^0$ K/min. Hall voltage measurements, as a function of temperature were carried out by van der Pauw method on rectangular bar samples in a cryogen free 15T magnet supplied by Cryogenic Ltd., UK.

## III. RESULTS AND DISCUSSIONS

XRD data of synthesized samples are depicted in Fig.1. All the diffraction patterns of the synthesized samples are crystallized in rhombohedral structure and indexed with space group $R\bar{3}m - D_{3d}^5$. Sb concentration dependent position of the highest intensity (015) peak and Full Width at Half Maxima (FWHM) of the corresponding peak have been estimated and given in Table I. Primarily position of this peak (015) shifts towards high angle side with increasing Sb concentration till 70% antimony content mixed crystal (Inset Fig.1).

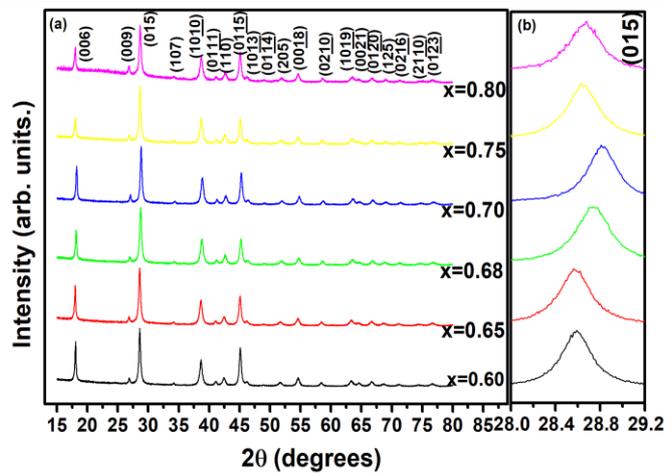

**Fig. 1.** (a) (color online) Room temperature x-ray powder diffraction patterns for the $(Bi_{1-x}Sb_x)_2Te_3$ (x=0.60, 0.65, 0.68, 0.70, 0.75 and 0.80) samples, synthesized by solid state reaction method. (b) 2θ shifting of (015) peak with Sb concentration.

It indicates decrease in crystal volume. It is obvious as replacement of Bi (atomic radii=1.60Å) with Sb (atomic radii=1.45Å) causes decrease in lattice parameter. After that, shift of peak position towards lower 2θ value may be related with defects, presence of elemental phases (unreacted/over stochiometric) and structural transition (discussed latter). K. Malik et al. also observed XRD peak position shift due to presence of elemental Bi/Sb in matrix of Bi-Sb alloy TE material [30]. Difference in vapour pressure between constituent elements causes segregation of Bi/Sb in Bi-Sb alloy [30]. However, anti-site defects at Bi (Sb), $Te^1$ and $Te^2$ sites occur frequently in $(Bi_{1-x}Sb_x)_2Te_3$. Structural and transport properties are strongly influenced by inherent Te vacancy ($V_{Te}$)



and/or $Te_{Bi}/Te_{Sb}$ (anti-site defects) in two end members, $Bi_2Te_3$ and $Sb_2Te_3$ [14, 31, 32]. FWHM may be related with strain in the unit cell, crystal quality, anisotropic growth of particle in samples [33]. FWHM increases with increasing Sb concentration in the mixed crystal till x=0.68. But, there is sudden drop in FWHM for $(Bi_{0.30}Sb_{0.70})_2Te_3$. Increasing Sb concentration in the mixed crystal causes transition from $Bi_2Te_3$ crystal configuration to $Sb_2Te_3$ crystal configuration. The $Te^1$ and $Te^2$ sites at different ambient environment causes $Sb-Te^2$ bond more polar or ionic than $Sb-Te^1$ [34]. The end member of the mixed crystal, $Sb_2Te_3$ shows p-type conductivity due to over stoichiometric Sb atoms along with native point defects and vacancies [31]. The over stoichiometric Sb atoms

TABLE I: Position of most intense x-ray diffraction peak (015), unit cell volume, $C_H/a_H$ ratio, FWHM of (015) and lattice strain value for different Sb concentration of $(Bi_{1-x}Sb_x)_2Te_3$ samples.

| Sb concentration (x) | Peak position (degree) | Unit cell volume ($Å^3$) | $C_H/a_H$ | FWHM (degree) | Strain($\varepsilon$) X $10^{-3}$ |
|---|---|---|---|---|---|
| 0.60 | 28.5745 | 524.3029 | 7.0746 | 0.3449 | 11.97 |
| 0.65 | 28.7392 | 522.3034 | 7.0830 | 0.3689 | 13.83 |
| 0.68 | 28.8214 | 521.1061 | 7.0870 | 0.3217 | 13.32 |
| 0.70 | 28.7400 | 520.3099 | 7.0945 | 0.3290 | 11.07 |
| 0.75 | 28.6450 | 518.3195 | 7.0959 | 0.3529 | 11.99 |
| 0.80 | 28.6620 | 516.3352 | 7.1090 | 0.3519 | 9.74 |

generally occupy $Te^2$ sites and give rise to $Sb_{Te}$ type anti-site defects [35]. However, solid solution $(Bi_{1-x}Sb_x)_2Te_3$ interestingly decreases defects due to increase in formation energy of defects. C. Niu et al. theoretically investigated most stable configuration of $(Bi_{1-x}Sb_x)_2Te_3$ for x=2/6 ~0.33 [8]. Hence, $(Bi_{0.70}Sb_{0.30})_2Te_3$ sample may be considered most stable structure amid synthesized polycrystalline samples. Segregation of minuscule quantity of constituent elements within polycrystalline mixed crystals (revealed by differential thermal analysis) are inherent owing to difference in vapour pressure. Anti-site defects, vacancy along with segregated elemental Bi/Sb/Te cause shift in diffraction peak.



However, at x=0.30 structural transition occurs towards end member, $Sb_2Te_3$ crystal configuration. This evolution is also accompanied with other physical properties viz., ρ-T, S-T and carrier concentration data, discussed latter.

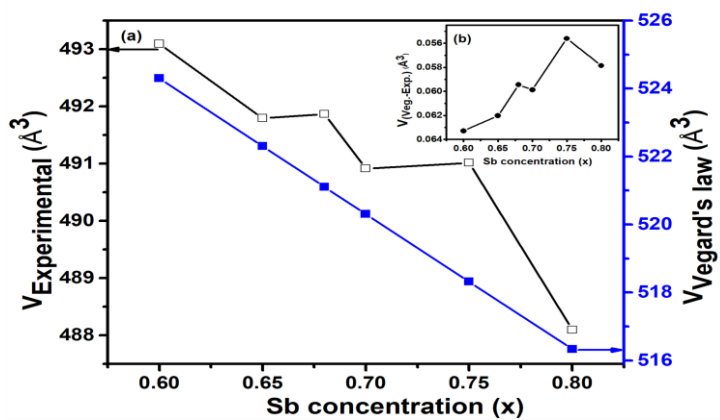

**Fig. 2.** (a) (color online) The unit cell volume of polycrystalline $(Bi_{1-x}Sb_x)_2Te_3$ (x=0.6, 0.65, 0.68, 0.70, 0.75 and 0.80) as a function of Sb concentration (x), estimated from Rietveld refinement (black filled) and Vegard's law (blue open). Inset (b) shows deviation of experimental from theoretical volume.

In depth structural analysis of the synthesized samples have been carried out by Rietveld refinement using MAUD software [36]. Experimentally obtained XRD patterns and theoretically fitted curve along with other parameters are given in supplementary information S1 [37]. Hexagonal unit cell volume of the synthesized mixed crystal is decreasing with increasing Sb concentration (Table I). Slight mismatch in lattice parameter of iso-structural $Sb_2Te_3$ and $Bi_2Te_3$ ascribe to develop alloy and follow Vegard's law, $a\{(Bi_{1-x}Sb_x)_2Te_3\}=\{a(1-x)\}Bi_2Te_3 + (ax)Sb_2Te_3$. Unit cell volume of the synthesized alloys decreases as lattice parameter of $Sb_2Te_3$ is less than $Bi_2Te_3$. However, little deviation from Vegard's law is observed due to increasing defect in $(Bi_{1-x}Sb_x)_2Te_3$ (0.60<x<0.80) (Fig. 2).

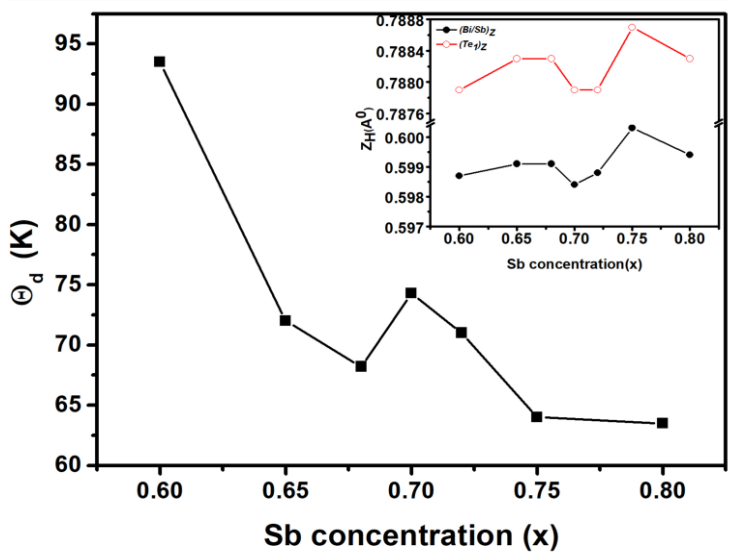

**Fig. 3.** (a) (color online) Debye temperatures ($θ_D$) as function of Sb-concentration (x) for $(Bi_{1-x}Sb_x)_2Te_3$ (x=0.60, 0.65, 0.68, 0.70, 0.75 and 0.80), estimated using equation (1). Inset shows positional disorder of Bi/Sb (black filled) and $Te_1$ (red open) as a function of Sb concentrations, obtained from Rietveld refinement of x-ray diffraction data.

Structural anisotropy, ratio of Hexagonal lattice parameter ($C_H/a_H$) increases with increasing Sb concentration (Table I). Deviation from Vegard's law increases, as given in Fig.2 inset may be due to increase in structural anisotropy. Further, Atomic vibration from their mean position i.e. Debye



Waller factor ($B_{iso}$) is also estimated from Rietveld refinement (Supplementary information S1) [37]. $B_{iso}$ increases with increasing Sb concentration (x), signify increase in positional disorder. It is noteworthy to mention that Sb-Te bond is weaker than Bi-Te bond, owing to difference in Pauling electronegativity [38]. As a result, mean square displacement increases with increasing Sb concentration. The large size difference between atomic radii of Bi (160pm) and Sb (145pm) may be another reason to occupy off centre position in quintuple layer and increase in $B_{iso}$ [36]. $B_{iso}$, within harmonic approximation is related to Debye temperature ($\theta_D$) by the following relation [39]:

$$B_{iso} = \left(\frac{6h^2}{Mk_B\theta_D}\right)\left[\frac{1}{4} + \left(\frac{T}{\theta_D}\right)^2 \int_0^{\frac{\theta}{T}} \frac{xdx}{e^x - 1}\right] \quad (1)$$

Estimated values of $\theta_D$ and corresponding positional disorders are given in Fig. 3 and inset shows the Z position of the atoms. It is noteworthy to mention that anomalous behaviours are also traced in $B_{iso}$, $Z_H$ and $\theta_D$ for x=0.70.

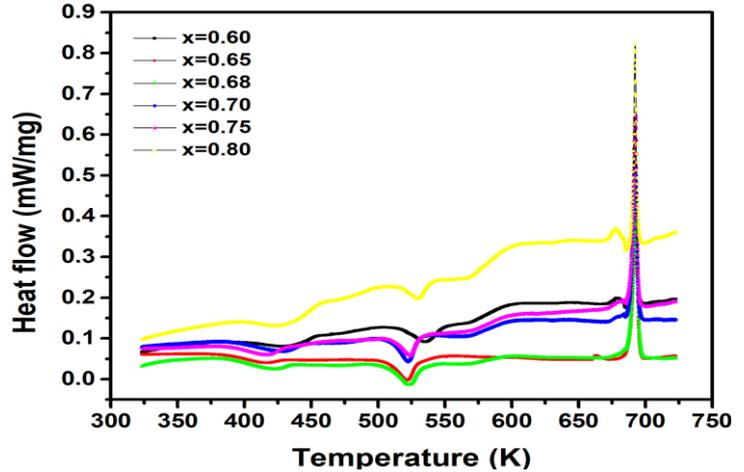

**Fig. 4.** (color online) DSC curve of polycrystalline $(Bi_{1-x}Sb_x)_2Te_3$ (x=0.60, 0.65, 0.68, 0.70, 0.75 and 0.80) at heating rate 10 °K/min.

DSC (differential scanning calorimetry) of the synthesized samples have been performed in inert gas atmosphere to avoid oxidation and presented in Fig. 4. DSC have been carried out in the temperature range 324 K to 722 K at heating rate 10 K/min. Endothermic peaks are observed near 420 K, 524 K in all DTA curves and comparatively smaller area is observed near 685 K. A broad exothermic peak at around 690K and small exothermic peak at 683K are observed. Areas of endothermic and exothermic peak have been estimated from the Fig. 4 and given in Table II. Area under the curve ($A_C$) signifies the enthalpy of the reaction. Hence, characteristic and amount of the reacting materials are directly related with $A_C$. The constituent elements and corresponding melting temperature of the mixed crystal



are Bi, Sb, Te: 544 K, 903 K and 722K respectively [30]. Endothermic peak at 420K may be related with temperature induced strain relaxation and inherent disorder within the crystal. It is noteworthy to mention that $A_C$ at peak temperature near 420K ($P_{T420}$) is strongly correlated with Sb concentration dependent strain value, obtained from powder XRD by employing W-H method (Table I) [Please find the supplementary information S2, for W-H plot] [37]. Bi and Sb are isostructural

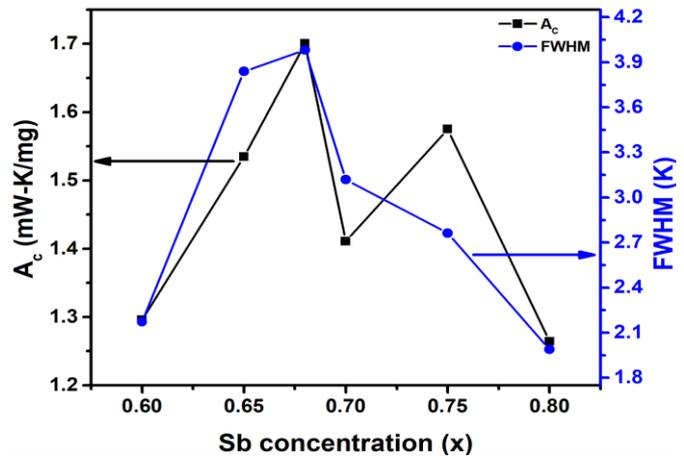

**Fig. 5.** (color online) Variation of area ($A_c$) under the exothermic peak and FWHM of exothermic peak near 692.42 K in DSC curve as a function of Sb concentration (x).

and isoelectronic. But, Pauling electronegativity of Bi, Sb and Te are 2.02, 2.05, 2.1 respectively [38]. Perceivable difference in electronegativity causes change in bond strength of Bi-Te and Sb-Te. The local structure become distorted and increases the disorder in the material [40, 41]. Strain within the mixed crystal decreases due to reduction of mass fluctuation in the unit cell of $(Bi_{1-x}Sb_x)_2Te_3$ with

**TABLE II.** Estimated values of Areas under endothermic, exothermic peak and $T_g$ from the DSC curve of $(Bi_{1-x}Sb_x)_2Te_3$ samples for different Sb concentration.

| Sb concentration (x) | Area under the DSC curve at particular position (mW/mgK) | | | | $T_g$ (K) |
|---|---|---|---|---|---|
| | Peak at 423.6226 K | Peak at 529.8745 K | Peak at 683.9156 K | Peak at 692.4195 K | |
| 0.60 | -0.60139 | -0.90580 | 0.16320 | 1.29606 | 569.50 |
| 0.65 | -0.43574 | -0.89410 | 0.05686 | 1.53472 | 575.79 |
| 0.68 | -0.59198 | -0.86397 | 0.05706 | 1.70048 | 568.13 |
| 0.70 | -0.73621 | -0.81843 | 0.11408 | 1.41089 | 567.18 |
| 0.75 | -0.56007 | -0.62659 | 0.15587 | 1.57519 | 568.52 |
| 0.80 | -1.47624 | -0.51430 | 0.37313 | 1.26443 | 567.95 |

increasing Sb concentration; but disorder increases. It is clearly reflected on the Sb concentration dependent strain and DTA analysis. It has been observed that the strain is maximum and disorder is minimum for x=0.65~0.68. Das et al. also reported that positional disorder has been affected due to bond strength in Se doped $Sb_2Te_3$ [23, 41]. However, anomalous change in strain and disorder has



been detected for x=0.70 in $(Bi_{1-x}Sb_x)_2Te_3$. This type of anomalous behaviour is also obtained in the structural analysis from powder XRD data (Fig.1 and Fig. 3). Endothermic peak near 524 K ($P_{T524}$) may be related with enthalpy for melting of minute amount of unreacted Bi in matrix of $(Bi_{1-x}Sb_x)_2Te_3$. Samples have been homogenised at 890 K for four days during synthesis. Saturation vapour pressures of Bi, Sb and Te at around 890K are 0.266 Pa, 10 Pa, 100 Pa respectively [42, 43]. Te vacancies ($V_{Te}$) in this pnictide chalcogenide are an inherent phenomenon due to higher volatilization of Te [32, 14]. Consequently, $(Bi_{1-x}Sb_x)_2Te_3$ samples always possess excesses of Bi/Sb. Increasing Sb concentration causes decrease of Bi in the matrix and concomitantly area of $P_{T524}$ decreases. Intense exothermic near 690K ($P_{T690}$) may be related with peak crystallization temperature ($T_P$) or reaction temperature of the mixed crystal. Bordering of the peak has been estimated by fitting with Lorentzian function. Fig. 5 shows variation of FWHM and $A_C$ of $P_{T690}$ with Sb concentration. Sharpness and small $A_C$ of $P_{T690}$ indicates perfection in crystal order along with decrease of other impurity within matrix of $(Bi_{1-x}Sb_x)_2Te_3$. Further, anomalous change in base line of DSC curve near $T_P$ signifies glass transition temperature ($T_g$) (Supplementary information S3) [37]. $T_g$, obtained from DSC curve are given in Table-II. $T_g$ of synthesized samples decreases with increasing Sb concentration due to formation of Sb-Te bond, weaker than Bi-Te bond. Further, $V_{Te}$ in mixed crystal increases for weak bond and high vapor pressure as explained above. It might result microscopic cluster of Te within crystal matrix. An iota of segregated Te is also supported by endothermic peak at 685K. The weak heteropolar bond and homopolar bond together leads to decrease in $T_g$ with increasing Sb in $(Bi_{1-x}SB_x)_2Te_3$. It is noteworthy to mention that $T_g$ also decreases in Sb-Sn-Te-Se based chalcogenide system owing to decrease in bond strength [38].

$\rho(T)$ curve of the synthesized samples have been depicted in Fig. 6. High temperature $\rho(T)$ data show activated behaviour, but slope of low temperature resistivity gradually reduces with decreasing temperature, after a certain temperature ($T_T$) (Table-III). The plateau after $T_T$ may be described by variable range hopping (VRH) as observed in $\rho(T)$ of TIs [44,45]. In general, activated behaviour of $\rho(T)$ is replaced by VRH at low temperature (T≤ 50K). Ren et al. shows that $\rho(T)$ of TI,



$Bi_{(2-x)}Sb_xTe_{(3-y)}Se_y$ solid solution may be fitted with equation: $\rho \propto \rho_0 exp\left[\left(\frac{T_0}{T}\right)^x\right]$ and obtained $x = 1/4$ [46]. It is noteworthy to mention that Topological insulating properties may be best viewed for compensated semiconductor [44]. Bi/Sb ratio in $(Bi_{1-x}Sb_x)_2Te_3$ mixed plays a crucial role to realize the TI properties by reducing bulk carrier concentration through reducing defects [8]. Most insulating state is obtained when Fermi level is in the midway of bulk conduction band and valence band. In this case, resistivity may be explained by theory of completely compensated semiconductor. Fluctuation in random potential due to defect and disorder causes bend in

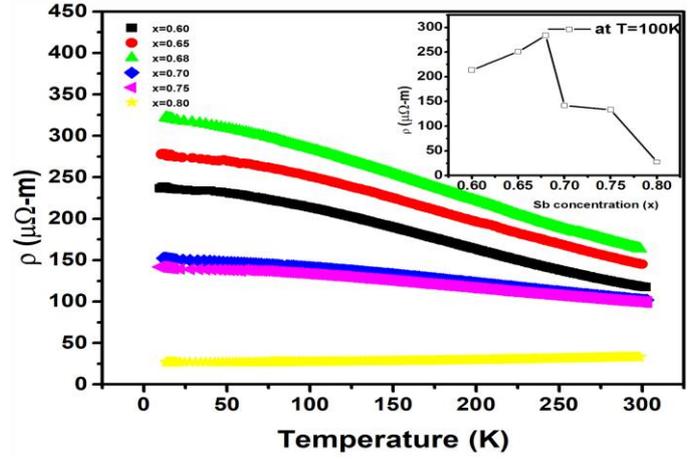

**Fig. 6. (a)**(color online) Thermal variation of Resistivity ($\rho$) for different Sb concentration (x) for polycrystalline $(Bi_{1-x}Sb_x)_2Te_3$ (x=0.60, 0.65, 0.68, 0.70, 0.75 and 0.80) samples. Inset shows resistivity ($\rho$) at typical temperature 100K for various Sb concentrations (x).

conduction and valence band near Fermi level. This large fluctuation is compensated or screened by creating electron and hole puddle in conduction or valence band [44]. At low temperature, $T_T \leq 50K$ electron or hole may tunnel between puddles and activated behaviour is replaced by VRH. Electrical transport in the VRH regime may be replaced by Efros-Shkolovskii law (ES) [45]:

$$\rho = \rho_0 exp\left(\frac{T_{ES}}{T}\right)^{0.5} \qquad (2)$$

i.e., $ln(\rho)$ Vs. $T^{-0.5}$ should show straight line behaviour below $T_T$, where, $\rho_0$ is temperature independent constant coefficient and $T_{ES}$ is related with the localization length of the energy states equivalent to Fermi Energy. Experimental and theoretically fitted curves are given in Supplementary information S4 [37]. However, Skinner et al. theoretically shows that activated conduction started above $T_T>50K$ [44]. It is noteworthy to mention that $\rho(T)$ of all the synthesized samples show VRH for $T_T\leq50K$ (See the supplementary information S4) [37]. Distinction between activated and VRH mechanism is subtle. However, $T_T$ is more prominent for comparatively compensated samples (discussed latter) viz., $(Bi_{0.32}Sb_{0.68})_2Te_3$, $(Bi_{0.35}Sb_{0.65})_2Te_3$ and $(Bi_{0.30}Sb_{0.70})_2Te_3$. Extended region of activated behaviour may



be related with presence of other conduction channel near Fermi surface due to inherent defects [46].

Non-monotonic ρ(T) with semiconducting nature are observed for Sb concentration (x) in between 0.60≤x≤0.70. Thermal gap of the semiconducting samples has been estimated using the equation:

$$\rho = \rho_0 exp\left(\frac{E_{act}}{k_B T}\right) \quad (3)$$

Where $E_{act}$, $k_B$ are activation energy and Boltzmann constant respectively. Estimated value of $\ln(\rho_0)$, $E_{act}$ and corresponding errors are given in Table-III. $\ln(\rho_0)$ decreases due to reduce of structural disorder and concomitantly carrier scattering followed by anomaly at x=0.70. Errors in $\ln(\rho_0)$ and $E_{act}$, as obtained during fitting using equation (3) has been shifted towards minimum value. It indicates spurious contribution due to disorder on the semiconducting behaviour decreases. However, ρ(T) of $(Bi_{0.20}Sb_{0.80})_2Te_3$ sample follows metallic nature. Metallic nature may arise due to mix of surface states and bulk states of the synthesized samples [36]. However, disorder and defects play a crucial role for metallic or semiconducting nature of ρ(T) data. In order to confirm origin of conductive nature for x=0.80; metallic curve has been fitted with power law equation (Please see the Supplementary information S5) [37]: $\rho = \rho_0 + AT^m$. According to Boltzmann transport mechanism, $\rho = \rho_{eph}T^m$; $\rho_{eph}$ indicates the electron-electron (e-e) or electron-phonon (e-ph) interaction. The best fit value, m=2 is obtained for the ρ(T) data, indicate the sample is purely metallic in nature

TABLE III. Estimated value of $\ln(\rho_0)$ and $E_{act}$ with their corresponding errors as obtained from fitted ρ(T) data using equation $\rho = \rho_0 e^{\left(\frac{E_{act}}{K_B T}\right)}$. Transition temperature ($T_T$) from VRH to activated behavior of ρ(T) data are also given.

| Sb concentration (x) | ln(ρ₀) | E_act (meV) | T_T (K) |
|---|---|---|---|
| 0.60 | 14.4052 ± 0.00513 | 39.1575 ± 0.2361 | 40.128 |
| 0.65 | 9.5686 ± 0.00692 | 38.1225 ± 0.3169 | 45.180 |
| 0.68 | 9.4078 ± 0.00817 | 35.8800 ± 0.3737 | 48.828 |
| 0.70 | 9.6241 ± 0.00417 | 22.5975 ± 0.1919 | 39.460 |
| 0.75 | 9.5816 ± 0.00300 | 18.9750 ± 0.0801 | 56.665 |
| 0.80 | - | - | 70.470 |



(Supplementary information S5 [37]). Metallic nature in end members of this mixed crystal may arise because of conducting bulk states due to excesses charge carriers originated by defects [47, 32, 16]. However, solid-solution $(Bi_{1-x}Sb_x)_2Te_3$ may reduce bulk carrier concentration by eliminating underlying defects in bulk material due to increase of formation energy of defects [8]. It nicely manifested in the $\rho(T)$ data for $0.60 \leq x \leq 0.75$ (fig. 6). Resistivity increases with increasing Sb concentration due to decrease in carrier concentration as well as increasing disorder in the mixed crystal. Further, it may be related with compensation of charge carrier with

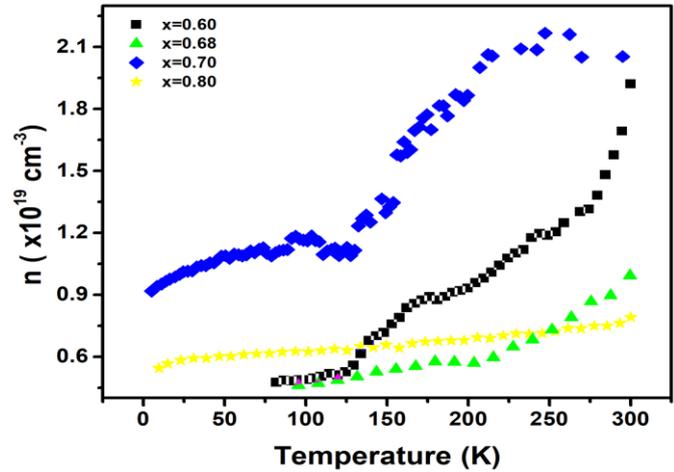

**Fig. 7.** (color online) Thermal variation of carrier concentration (n) for different Sb concentration(x) of polycrystalline $(Bi_{1-x}Sb_x)_2Te_3$ samples (x=0.60, 0.68, 0.70 and 0.80).

increasing Sb concentration till x=0.68. However, unanticipated drop in $\rho(T)$ have been observed after x=0.70. It may be related with increase in crystalline order and carrier concentration (n) in $(Bi_{1-x}Sb_x)_2Te_3$. In order to confirm about n; Hall measurements have been carried out for four samples viz., x=0.60, 0.68, 0.70, and 0.80 (Fig.7). Carrier concentration due to defect, dominate for x=0.80 Sb contained sample and shows metallic nature. Hence, comparatively minimum variation of n with temperature is obtained for $(Bi_{0.20}Sb_{0.80})_2Te_3$ sample and correlated with the metallic nature of

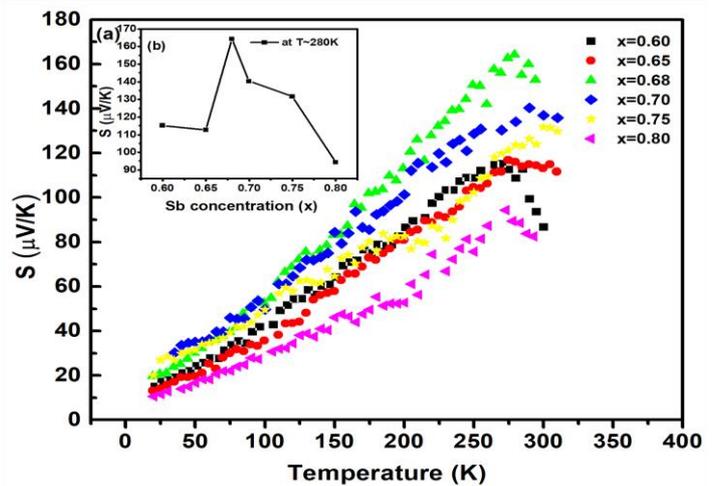

**Fig. 8.** (color online) (a)Temperature dependent thermopower (S) for polycrystalline $(Bi_{1-x}Sb_x)_2Te_3$ samples (x=0.60, 0.65, 0.68, 0.70, 0.75 and 0.80). Inset (b) shows S at typical temperature 270K for different Sb concentration (x) and anomaly is observed at x=0.70.

$\rho(T)$. However, rapid decrease in n bellow 150K may be related with the plateau observed at low temperature $\rho(T)$. It also signifies the transition of $\rho(T)$ from activated to VRH mechanism.



Distinction between VRH and activated mechanism is a subtle issue and depends on degree of disorder within the samples [46]. Further, increase in density of state effective mass after T~150K (Given below) may be related with creation of puddle in conduction and valence band. It is noteworthy to mention that puddle in conduction and valence band are sole responsible for VRH mechanism [43]. However, ρ(T) with varying Sb concentration are corroborated with n(T) data and thermal variation of (n) also correlated with semiconducting (for x=0.60, 0.68 and 0.70) and metallic (for x=0.80) characteristic of synthesized samples.

Fig. 8 represents thermal variation of thermopower (S(T)) down to 20K. S(T) increases with temperature, indicates degenerate semiconducting nature of the synthesized samples. It is interesting to note that Sb concentration dependent S(T) value is correlated with resistivity and other estimated quantity. Disorder within mixed crystal matrix strongly affect carrier conductivity in temperature gradient and concomitantly S(T). Disorder to comparatively ordered transition at x~0.70 causes sudden drop in S(T) value (Fig.8 inset). It further supports disorder induced iso-structural phase transition for x~0.70. Highest S value is obtained for most compensated $(Bi_{0.32}Sb_{0.68})_2Te_3$ followed by the anomaly at x=0.70. S(T) depict positive value, indicate p-type nature of the samples and also supported by estimated n (Fig. 7) using Hall voltage. It is noteworthy to mention that defects and disorder act as scattering centre and strongly influence S(T). Total S value may be considered as sum of diffusion thermopower and phonon interaction term (e-ph). In order to estimate effect of scattering mechanism, S(T) have been fitted with following equation in Debye temperature limit [48]:

$$S(T) = AT + BT^3 \qquad (4)$$

Here, first and second terms are contribution of diffusion and e-ph interaction in S. The estimated value of A ($= \frac{\pi^2 k_B^2}{3qE_F}$), B and corresponding values of $E_F$ are listed in Table IV. Both, A and B increases with Sb concentration till x=0.68, indicates increase in scattering due to disorder. It should be mentioned that $E_F$ is related with optical band gap ($E_g$) of the synthesized material rather than the activation energy. Band structure of end members have been studied extensively [49, 50, 14]. However, exact band gap ($E_g$) of the materials is still illusive. Whereas, Smith et al. reported that



degeneracy corrected $E_g$=0.2eV for 80% $Sb_2Te_3$ contained $Bi_2Te_3$-$Sb_2Te_3$ mixed crystal [51]. Michiardi et al. theoretically and experimentally estimated $E_g$~120 eV for single crystal $Bi_2Te_3$ [50]. Synthesized samples show p-type conductivity and ρ(T), S(T) follow degenerate semiconducting nature. Hence, position of Fermi surface (or $E_F$) is near the valence band [52]. Compensation of charge carriers in the samples cause shift in $E_F$. Population of transport carrier have been enhanced due to inherent $V_{Te}$ causes shift in $E_F$ towards valence band as estimated from A coefficients. e-ph scattering plays an important role in transport properties with increasing temperature. Phonon scattering in various TE materials have been also revealed [53]. It is noteworthy to mention that phonon scattering may also help in enhancement of ZT by reducing κ of the material. However, there are very limited reports in correlation of structural and electronic properties with A and B parameter. In this report, disorder induced scattering and defect induced carrier concentration are nicely depicted in estimated A, B coefficients.

TABLE IV. The best fit value of A, B and $E_F$ parameters as obtained from fitted thermopower data employing equation S= A T + B T$^3$ and room temperature lattice thermal conductivity, estimated using equation, $\kappa_l = \Lambda \frac{\bar{M}\theta_D^3 \delta}{\gamma_g^2 m^{\frac{2}{3}} T}$ for $(Bi_{1-x}Sb_x)_2Te_3$ samples.

| Sb concentration(x) | A (μV/K$^2$) | B (μV/K$^4$) | $E_F$ (eV) | $\kappa_l$ (W/mK) |
|---|---|---|---|---|
| 0.60 | 0.3762 ± 0.0155 | 6.9275E-7 ± 3.4533E-7 | 0.06488 | 0.82578 |
| 0.65 | 0.2868 ± 0.0156 | 2.5694E-6 ± 3.5597E-7 | 0.06833 | 0.37202 |
| 0.68 | 0.4346 ± 0.0245 | 3.0519E-6 ± 7.2050E-7 | 0.05478 | 0.31378 |
| 0.70 | 0.4296 ± 0.0492 | 7.8484E-7 ± 1.1585E-7 | 0.08111 | 0.40350 |
| 0.75 | 0.3537 ± 0.0222 | 1.6040E-7 ± 5.2522E-7 | 0.07948 | 0.25452 |
| 0.80 | 0.2309 ± 0.0160 | 1.0113E-6 ± 4.3196E-7 | 0.08918 | 0.24487 |

It is evident from the ρ(T) and S(T) data that synthesized samples are degenerate in nature. Underlying defects and $V_{Te}$ have been made samples potential extrinsic semiconductor. However, these degenerate extrinsic semiconductors may also be expressed by the following Pisarenko relation [54]:



$$S = \frac{8\pi^2 k_B^2 T}{3eh^2} m^* \left(\frac{\pi}{3n}\right)^{2/3} \tag{5}$$

where, $k_B$, e, h, $m^*$, n are Boltzman constant, charge of the carrier, Plank constant, density of state effective mass (DOS) and carrier concentration. Temperature dependent DOS effective mass for synthesized samples have been estimated from equation 5 and presented in Fig. 9. The estimated values are in good agreement with previously reported result at room temperature [55]. DOS effective mass increases in $(Bi_{1-x}Sb_x)_2Te_3$ with

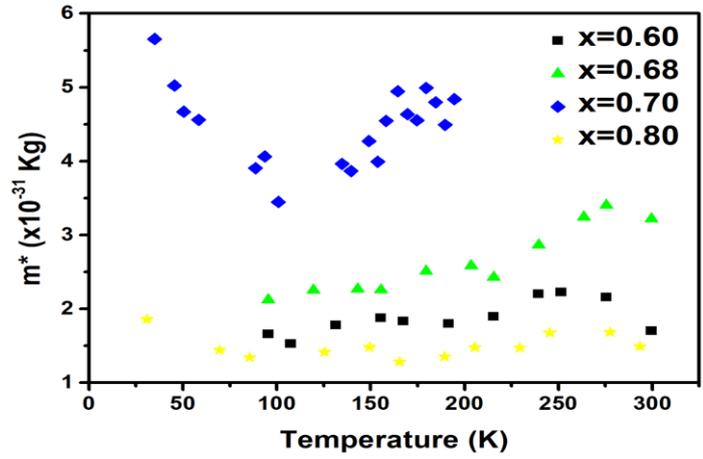

**Fig. 9.** (color online) Thermal variation of Density of state effective mass ($m^*$) for different Sb concentration(x) of polycrystalline $(Bi_{1-x}Sb_x)_2Te_3$ (x=0.60, 0.68, 0.70 and 0.80).

increasing Sb concentration but decreases with decreasing temperature. Incremental behaviour at low temperature may be related with carrier transport by VRH mechanism as observed in ρ(T) curve. DOS effective mass decreases near room temperature due to complex interplay of different degenerate levels. Further, ZT have been estimated using S, ρ and $\kappa_L$. According to Slack, $\kappa_L$ may be expressed by bellow equation in Umklapp scattering limit [56]

$$\kappa_L = \Lambda \frac{\bar{M}\, \theta_D^3 \delta}{\gamma_g^2 m^{2/3} T} \tag{6}$$

Where $\bar{M}, \delta^3, m$ represent average atomic mass in unit cell, volume per atom, and number of atoms in rhombohedral unit cell. Here we have estimated $\kappa_L$ considering $\gamma_G$, Gruneisen parameter is equal to value of $Sb_2Te_3$. It should be noted that $\gamma_G$ is almost temperature independent, but depend on stoichiometry of constituent elements. In this study we deal with a small variation of Sb in $(Bi_{1-x}Sb_x)_2Te_3$. In this context $\gamma_G$ of $Sb_2Te_3$ has been taken 1.7(1) [57]. However, estimated $\kappa_L$ are given in Table 4 and room temperature ZT is presented in Fig.10. $\kappa_L$ decreases with antimony concentration and structural transition at x=0.70 also corroborates with fluctuation in ZT value near x=0.70 in Fig.10. Sb concentration dependent ZT changes slope at $(Bi_{0.30}Sb_{0.70})_2Te_3$ samples. Chemical bond of Bi-Te is stronger than Sb-Te bond. Weak chemical bond decreases the sound velocity in the mixed crystal. Further, Structural anisotropy increases with increasing Sb concentration. Das et al. observed correlation between structural anisotropy and phonon anharmonicity for $Sb_2Te_{3-x}Se_x$ [58]. Further, local structural disorder due to off centring of atoms and deformation due to $V_{Te}$ in unit cell increases. These mismatch between neighbour causes reflection of phonon [59]. These might be plausible explanations for the decrease in $\kappa_L$.

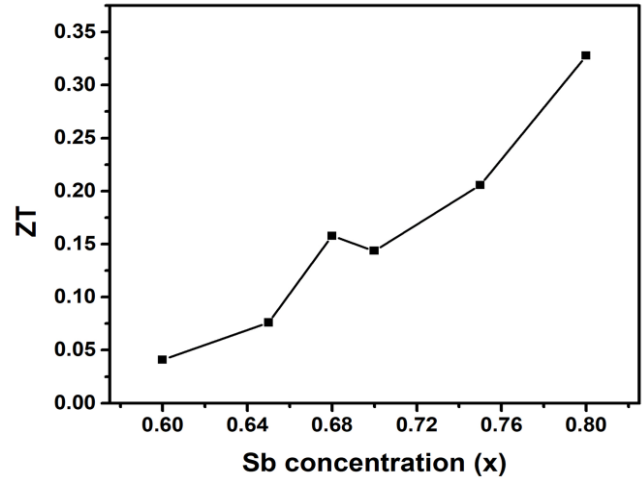

**Fig. 10.** Sb concentration dependent Figure of merit (ZT) of polycrystalline $(Bi_{1-x}Sb_x)_2Te_3$ (x=0.6, 0.65, 0.68, 0.70, 0.75 and 0.80) at room temperature.

## IV. CONCLUSION

Positional disorder and defects strongly influence the TE properties of $(Bi_{1-x}Sb_x)_2Te_3$ (x=0.60, 0.65, 0.68, 0.70 and 0.80) alloy. In depth structural analysis reveals that most stable and order configuration is $(Bi_{0.30}Sb_{0.70})_2Te_3$. Disorder induced iso-structural phase transition occurs for x=0.70 and anomaly observed on other structural parameters viz., Rhombohedral angle, volume, atomic position and estimated Debye temperature. Sb concentration dependent atomic disorder and anisotropic vibration at mean atomic position indicates similar anomaly. DSC measurement confirms glass transition ability of the mixed crystal and segregation of an iota of elemental material within the matrix. Inherent disorder and defects strongly influenced the S(T) and ρ(T) data. Low temperature ρ(T) follows VRH type conductivity whereas high temperature depicts semiconducting nature for $0.60 \leq x \leq 0.75$. $(Bi_{0.20}Sb_{0.80})_2Te_3$ is metallic owing to presence of defect and vacancy induced excess charge carrier. Estimated $E_{act}$ is strongly correlated with disorder induced structural anomaly. Further, compensation of charge carrier and prominence of VRH behaviour strongly depends on Sb/Bi ratio and correlated with ρ(T) and n data. ZT of the synthesized samples have been estimated. The structural and transport properties of the synthesized samples are corroborated. It may be concluded that disorder to ordered iso-structural phase transition occurs at x=0.70 in synthesized samples. Synthesized mixed crystals have glass transition ability near 578 K. It is noteworthy to mention that degenerate semiconducting nature and excess charge carrier aid to enhance TE properties of $(Bi_{1-x}Sb_x)_2Te_3$ mixed crystal.

**SUPPLEMENTARY MATERIAL**

Supplementary information for the details of x-ray diffraction patterns after Rietveld refinement and the corresponding refinement parameters obtained using MAUD software, estimation of grain size and strain from Willamson-Hall plot, evaluation of Glass transition temperature, fitting of law temperature resistivity data using Variable range hopping mechanism (VRH) for polycrystalline $(Bi_{(1-x)}Sb_x)_2Te_3$ (x = 0.60, 0.65, 0.68, 0.70, 0.75 and 0.80 ) samples and fitting using power law equation of $\rho(T)$ data of a typical polycrystalline $(Bi_{0.2}Sb_{0.8})_2Te_3$ sample.

**AUTHORS' CONTRIBUTIONS**

Contributions of authors are according to the author list.

**ACKNOWLEDGEMENTS**

This work is supported by Science and Engineering Research Board (SERB), Govt. of India in the form of sanctioning research project, File Number: EEQ/2018/001224. Author SM is thankful to CSIR, India for providing Research Fellowships.

**DATA AVAILABLITY**

The data that support the findings of this study are available from the corresponding author upon reasonable request.

# Supplemental Online Material

**FIGURE S1:** X-ray diffraction patterns after Rietveld refinement pattern and the corresponding refinement parameters obtained using MAUD software for $(Bi_{1-x}Sb_x)_2Te_3$ with x = 0.60, 0.65, 0.68, 0.70, 0.75 and 0.80 samples at room temperature.

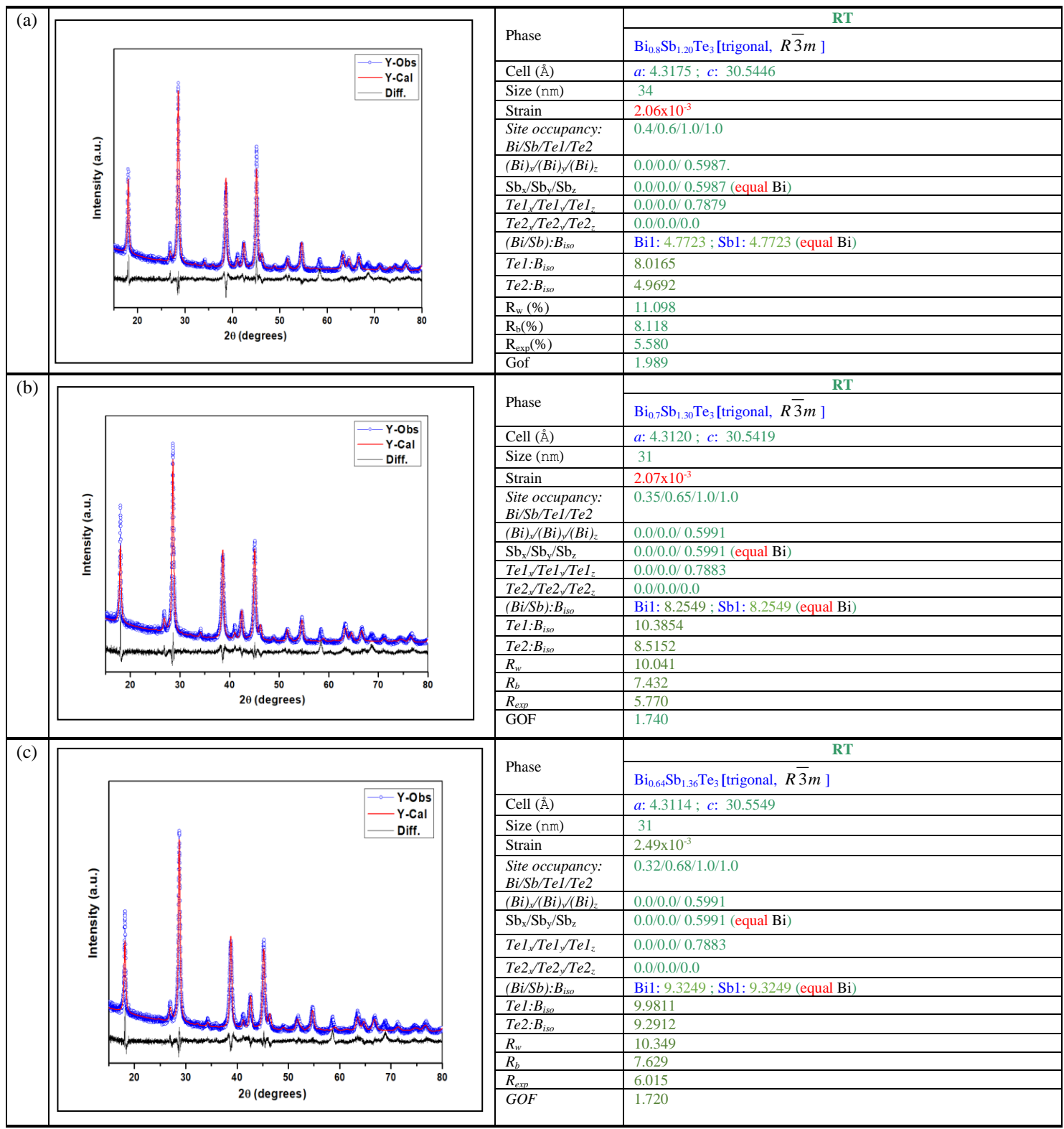

(a)

| Phase | RT |
| --- | --- |
| | $Bi_{0.8}Sb_{1.20}Te_3$ [trigonal, $R\bar{3}m$] |
| Cell (Å) | a: 4.3175 ; c: 30.5446 |
| Size (nm) | 34 |
| Strain | $2.06 \times 10^{-3}$ |
| Site occupancy: Bi/Sb/Te1/Te2 | 0.4/0.6/1.0/1.0 |
| $(Bi)_x/(Bi)_y/(Bi)_z$ | 0.0/0.0/ 0.5987. |
| $Sb_x/Sb_y/Sb_z$ | 0.0/0.0/ 0.5987 (equal Bi) |
| $Te1_x/Te1_y/Te1_z$ | 0.0/0.0/ 0.7879 |
| $Te2_x/Te2_y/Te2_z$ | 0.0/0.0/0.0 |
| $(Bi/Sb):B_{iso}$ | Bi1: 4.7723 ; Sb1: 4.7723 (equal Bi) |
| $Te1:B_{iso}$ | 8.0165 |
| $Te2:B_{iso}$ | 4.9692 |
| $R_w$ (%) | 11.098 |
| $R_b$(%) | 8.118 |
| $R_{exp}$(%) | 5.580 |
| Gof | 1.989 |

(b)

| Phase | RT |
| --- | --- |
| | $Bi_{0.7}Sb_{1.30}Te_3$ [trigonal, $R\bar{3}m$] |
| Cell (Å) | a: 4.3120 ; c: 30.5419 |
| Size (nm) | 31 |
| Strain | $2.07 \times 10^{-3}$ |
| Site occupancy: Bi/Sb/Te1/Te2 | 0.35/0.65/1.0/1.0 |
| $(Bi)_x/(Bi)_y/(Bi)_z$ | 0.0/0.0/ 0.5991 |
| $Sb_x/Sb_y/Sb_z$ | 0.0/0.0/ 0.5991 (equal Bi) |
| $Te1_x/Te1_y/Te1_z$ | 0.0/0.0/ 0.7883 |
| $Te2_x/Te2_y/Te2_z$ | 0.0/0.0/0.0 |
| $(Bi/Sb):B_{iso}$ | Bi1: 8.2549 ; Sb1: 8.2549 (equal Bi) |
| $Te1:B_{iso}$ | 10.3854 |
| $Te2:B_{iso}$ | 8.5152 |
| $R_w$ | 10.041 |
| $R_b$ | 7.432 |
| $R_{exp}$ | 5.770 |
| GOF | 1.740 |

(c)

| Phase | RT |
| --- | --- |
| | $Bi_{0.64}Sb_{1.36}Te_3$ [trigonal, $R\bar{3}m$] |
| Cell (Å) | a: 4.3114 ; c: 30.5549 |
| Size (nm) | 31 |
| Strain | $2.49 \times 10^{-3}$ |
| Site occupancy: Bi/Sb/Te1/Te2 | 0.32/0.68/1.0/1.0 |
| $(Bi)_x/(Bi)_y/(Bi)_z$ | 0.0/0.0/ 0.5991 |
| $Sb_x/Sb_y/Sb_z$ | 0.0/0.0/ 0.5991 (equal Bi) |
| $Te1_x/Te1_y/Te1_z$ | 0.0/0.0/ 0.7883 |
| $Te2_x/Te2_y/Te2_z$ | 0.0/0.0/0.0 |
| $(Bi/Sb):B_{iso}$ | Bi1: 9.3249 ; Sb1: 9.3249 (equal Bi) |
| $Te1:B_{iso}$ | 9.9811 |
| $Te2:B_{iso}$ | 9.2912 |
| $R_w$ | 10.349 |
| $R_b$ | 7.629 |
| $R_{exp}$ | 6.015 |
| GOF | 1.720 |

(d)
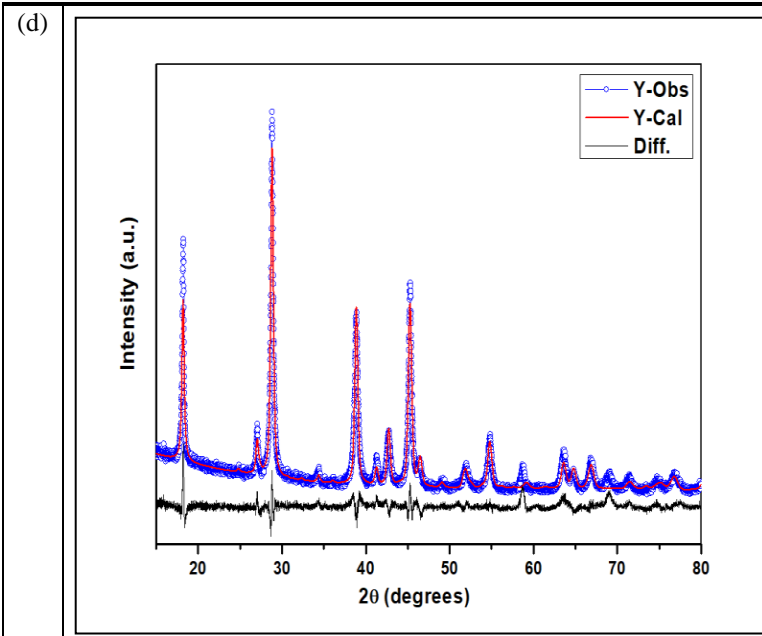

| | RT |
|---|---|
| Phase | $Bi_{0.6}Sb_{1.40}Te_3$ [trigonal, $R\bar{3}m$] |
| Cell (Å) | $a$: 4.3088 ; $c$: 30.5566 |
| Size (nm) | 35 |
| Strain | $2.32 \times 10^{-3}$ |
| Site occupancy: Bi/Sb/Te1/Te2 | 0.3/0.7/1.0/1.0 |
| $(Bi)_x/(Bi)_y/(Bi)_z$ | 0.0/0.0/ 0.5984 |
| $Sb_x/Sb_y/Sb_z$ | 0.0/0.0/ 0.5984 (equal Bi) |
| $Te1_x/Te1_y/Te1_z$ | 0.0/0.0/ 0.7879 |
| $Te2_x/Te2_y/Te2_z$ | 0.0/0.0/0.0 |
| $(Bi/Sb):B_{iso}$ | Bi1: 7.8518 ; Sb1: 7.8518 (equal Bi) |
| $Te1:B_{iso}$ | 9.6399 |
| $Te2:B_{iso}$ | 7.3381 |
| $R_w$ | 10.777 |
| $R_b$ | 7.964 |
| $R_{exp}$ | 5.978 |
| GOF | 1.803 |

(e)
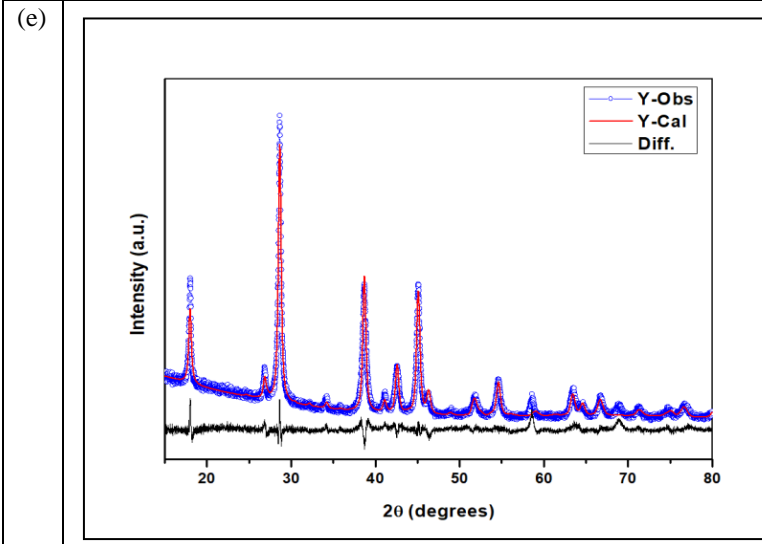

| | RT |
|---|---|
| Phase | $Bi_{0.5}Sb_{1.5}Te_3$ [trigonal, $R\bar{3}m$] |
| Cell (Å) | $a$: 4.3071; $c$: 30.5629 |
| Size (nm) | 30 |
| Strain | $1.83 \times 10^{-3}$ |
| Site occupancy: Bi/Sb/Te1/Te2 | 0.25/0.75/1.0/1.0 |
| $(Bi)_x/(Bi)_y/(Bi)_z$ | 0.0/0.0/ 0.6003 |
| $Sb_x/Sb_y/Sb_z$ | 0.0/0.0/ 0.6003 (equal Bi) |
| $Te1_x/Te1_y/Te1_z$ | 0.0/0.0/ 0.7887 |
| $Te2_x/Te2_y/Te2_z$ | 0.0/0.0/0.0 |
| $(Bi/Sb):B_{iso}$ | Bi1: 10.8614 ; Sb1: 10.8614 (equal Bi) |
| $Te1:B_{iso}$ | 9.4759 |
| $Te2:B_{iso}$ | 10.2802 |
| $R_w$ | 9.969 |
| $R_b$ | 7.208 |
| $R_{exp}$ | 5.859 |
| GOF | 1.701 |

(f)
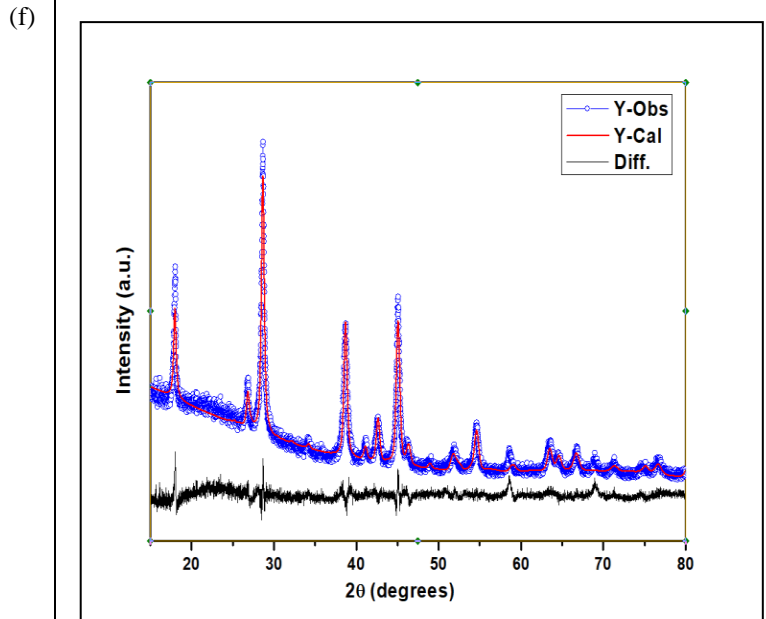

| | RT |
|---|---|
| Phase | $Bi_{0.4}Sb_{1.6}Te_3$ [trigonal, $R\bar{3}m$] |
| Cell (Å) | $a$: 4.2959; $c$: 30.5397 |
| Size (nm) | 32 |
| Strain | $2.45 \times 10^{-3}$ |
| Site occupancy: Bi/Sb/Te1/Te2 | 0.2/0.8/1.0/1.0 |
| $(Bi)_x/(Bi)_y/(Bi)_z$ | 0.0/0.0/ 0.5994 |
| $Sb_x/Sb_y/Sb_z$ | 0.0/0.0/ 0.5994 (equal Bi) |
| $Te1_x/Te1_y/Te1_z$ | 0.0/0.0/ 0.7883 |
| $Te2_x/Te2_y/Te2_z$ | 0.0/0.0/0.0 |
| $(Bi/Sb):B_{iso}$ | Bi1: 11.1399 ; Sb1: 11.1399 (equal Bi) |
| $Te1:B_{iso}$ | 8.5022 |
| $Te2:B_{iso}$ | 8.0360 |
| $R_w$ | 8.697 |
| $R_b$ | 6.730 |
| $R_{exp}$ | 6.001 |
| GOF | 1.449 |

**FIGURE S2:** Williamson-Hall plot of polycrystalline $(Bi_{1-x}Sb_x)_2Te_3$ (x = 0.60, 0.65, 0.68, 0.70, 0.75 and 0.80). Solid line represents best linear fitted line.

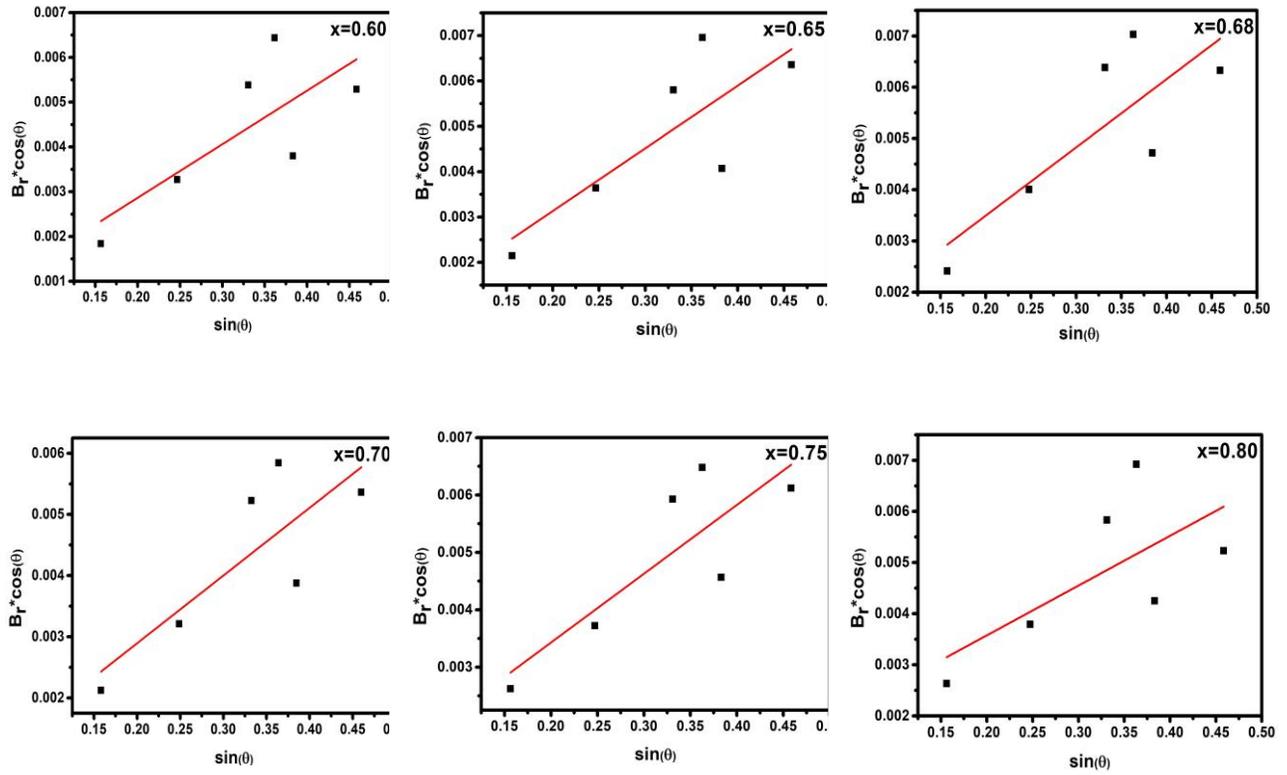

**FIGURE S3:** Glass transition temperature ($T_g$) of polycrystalline $(Bi_{1-x}Sb_x)_2Te_3$ (x = 0.60, 0.65, 0.68, 0.70, 0.75 and 0.80 ) samples from DSC curve. Temperature corresponding to the intersection of two linear fitted line is glass transition temperature.

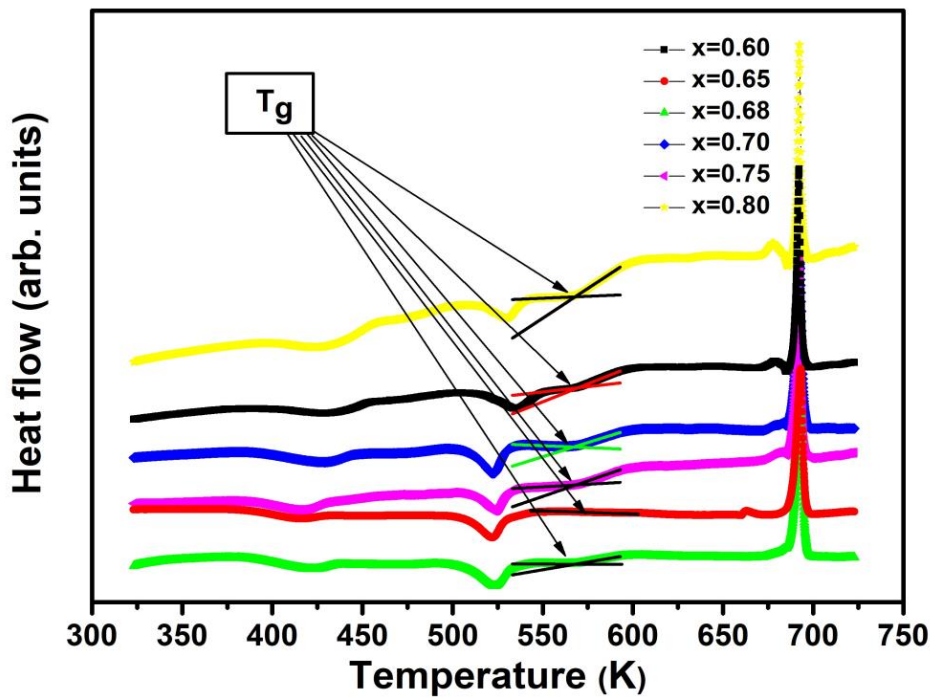

**FIGURE S4:** Low temperature $\rho(T)$ data are fitted by employing Variable-range-hopping (VRH) mechanism; $\ln(\rho)$ vs $T^{-1/2}$ data are plotted for polycrystalline $(Bi_{1-x}Sb_x)_2Te_3$ Samples (x = 0.60, 0.65, 0.68, 0.70, 0.75 and 0.80). Sloid lines represent best linear fitted curve to VRH regions.

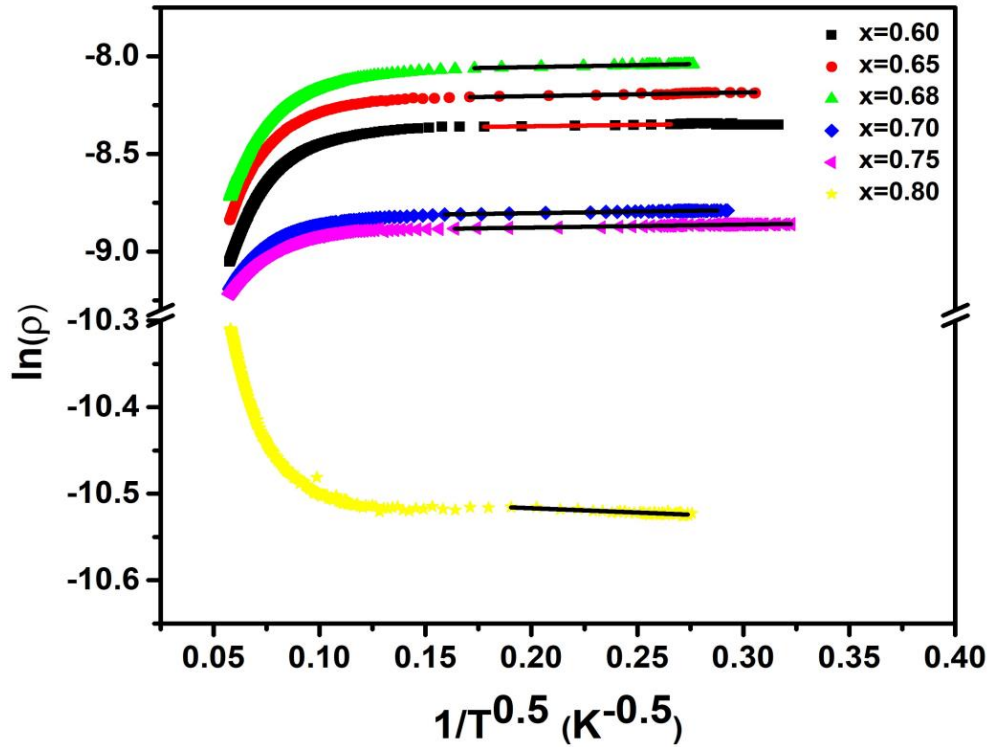

**FIGURE S5:** The typical sample $(Bi_{0.20}Sb_{0.80})_2Te_3$ is fitted with power law equation, $\rho = \rho_0 + AT^m$. Solid line represents best fitted curve. m=2 is obtained from best fit, indicates the sample is purely metallic in nature.

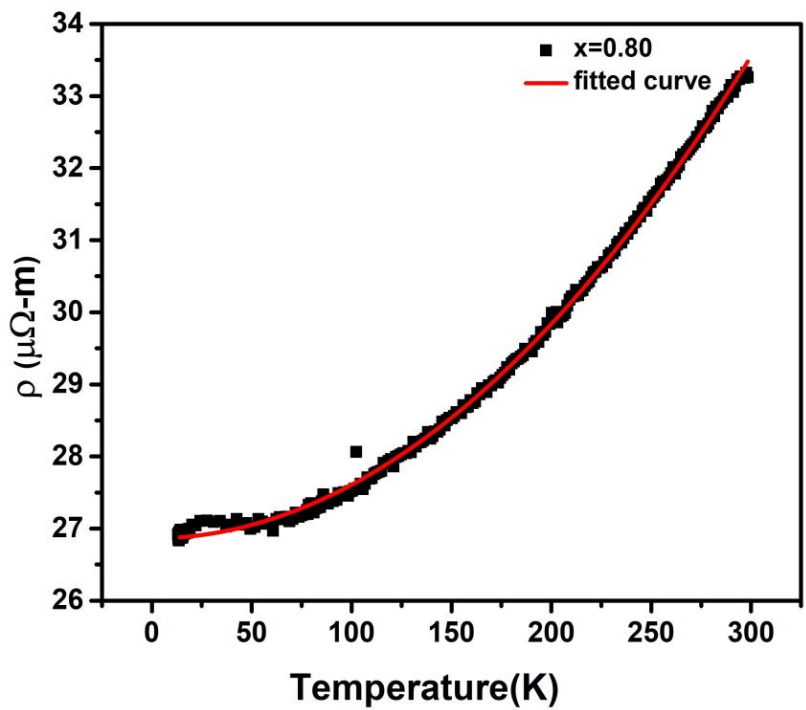